\documentclass[usenatbib,letter]{mnras}

\usepackage{graphicx,rotating}
\usepackage{amssymb,amsmath,pdflscape}
\usepackage{natbib}
\usepackage{notoccite}

\newcommand{\hphi}{\hat{\Phi}}

\newcommand{\rc}{R_c}

\newcommand{\hbr}{\hat{R}}
\newcommand{\hbz}{\hat{Z}}

\begin{document}

\title[Rigidly rotating spheroid-ring systems]{Rigidly rotating, incompressible spheroid-ring systems: new bifurcations, critical rotations and degenerate states}

\author[B. Basillais and J.-M. Hur\'e]
{B. Basillais$^{1,2}$\thanks{E-mail:baptiste.boutin-basillais@u-bordeaux.fr},
and J.-M. Hur\'e$^{1,2}$\\
$^{1}$Univ. Bordeaux, LAB, UMR 5804, F-33615, Pessac, France\\
$^{2}$CNRS, LAB, UMR 5804, F-33615, Pessac, France}

\date{Received ??? / Accepted ???}
 
\pagerange{\pageref{firstpage}--\pageref{lastpage}} \pubyear{???}

\maketitle

\label{firstpage}

\begin{abstract}
The equilibrium of incompressible spheroid-ring systems in rigid rotation is investigated by numerical means for a unity density contrast. A great diversity of binary configurations is obtained, with no limit neither in the mass ratio, nor in the orbital separation. We found only detached binaries, meaning that the end-point of the $\epsilon_2$-sequence is the single binary state in strict contact, easily prone to mass-exchange. The solutions show a remarkable confinement in the rotation frequency-angular momentum diagram, with a total absence of equilibrium for $\Omega^2/ \pi G \rho \gtrsim 0.21$. A short band of degeneracy is present next to the one-ring sequence. We unveil a continuum of bifurcations all along the ascending side of the Maclaurin sequence for eccentricities of the ellipsoid less than $\approx 0.612$ and which involves a gradually expanding, initially massless loop.
\end{abstract}

\begin{keywords}
gravitation | methods: numerical | stars: interiors | (stars:) binaries: general | stars: rotation 
\end{keywords}

\section{Introduction}

Despite its academic character, the theory of figures gives essential information on self-gravitating fluids like the mass, size, shape, rotation rate and energy content, which can be confronted with observations \citep{horedttextbook2004}. A broad range of applications is concerned, e.g. normal and compact stars, binaries, interstellar cores, planets (and initially the Earth), and galaxies \citep[see e.g.][]{chandra69}. There is a large variety of allowed configurations depending on the equation of state, prescribed rotation profile, internal circulations and environmental effects like external gravity, magnetic fields or ambient pressure \citep[see e.g.][]{hachisu86,fe14,hhn18}. The possibility of discovering new states, even in the incompressible and axisymmetrical limits, is an exciting source of motivation and also very challenging from a technical point of view \citep[][]{he84,henii86,nel92,wst92}.

In a pioneering paper, \cite{erihach83crrr} have studied the conditions for ``core-ring'' and especially ``ring-ring'' equilibria and their position in the classical $\omega^2$-$j^2$ diagram. In particular, they showed that there is a maximum rotation rate for core-ring states in rigid rotation. More recently, \cite{ansorg03} have characterized the bifurcations along the descending part of the Maclaurin sequence (MLS) with an unprecedented accuracy from spectral decomposition of the fluid boundary \citep[see also][]{hen86}. They computed new series of figures, denoted $\epsilon_{k \ge 2}$, which display a strong flattening and a marked wavy shape. When $k$ is an even number, $\frac{k}{2}$ pinchings are present on both sides of the equatorial plane. At the end-point of the corresponding sequence, the spheroid is on the verge of splitting into a central core surrounded by $m-1=\frac{k}{2}$ ring(s). A similar phenomenon exists when $k$ is an odd number (a single ring may lead to multiple, concentric ringlets). We have continued this investigation for $k=2$ (i.e., the ``spheroid-ring'' case) to determine which states can really exist beyond the sequence ending. This kind of configuration is important for the physics of accretion disks and tori around normal and compact stars \citep{mne98,akl98}, mass-transfer, rotation and angular momentum exchange between the two components. For this purpose, we have used a new version of the {\tt DROP} code \citep{hh17} which enables to take into account $m>1$ detached bodies in mutual interaction. The wide exploration of the parameter space permits to answer several major pending questions. In particular, no contact binary other than the one reported in \cite{ansorg03} was detected. The equilibrium states clearly fill the bottom part of the $\omega^2$-$j^2$ diagram (the rotation frequency must not exceed a certain threshold), while they slightly overtake the one-ring sequence (ORS) \citep[e.g.][]{hachisu86}. In some cases, two different states are characterized by the same rotation rate and angular momentum. Besides, we find a collection of new routes linking the MLS and the ORS for ellipsoids with an initial eccentricity lower than $\approx 0.612$. These results, once expanded to compressible systems, might have interesting implications in various domains, e.g. the physics of fast rotators and transient rings around Be-stars \citep{mszk06}, accretion around compact stars \citep{audrey18}, the torus in active galactic nuclei \citep{mas15}, massive disks and rings orbiting proto-stars \citep{km06,ti16}, planetary rings \citep{wt88}, and the formation of the Earth-Moon system	 \citep{ls17}.

\section{Theoretical background}

A figure of equilibrium basically obeys a first integral for the fluid motion and thermodynamical content $H - \Psi - \Phi=$const., which combines the enthalpy $H=\int{dP/\rho}$ of the fluid, the gravitational potential $\Psi$ and the centrifugal potential $\Phi$ \citep{horedttextbook2004}. As for any self-gravitating fluid, $\Psi$ and $\rho$ are linked through the Poisson equation $\Delta \Psi = 4 \pi G \rho$. In the incompressible limit of interest here, the potential can be expressed as a one-dimensional integral over the fluid boundary \citep{wong74,ansorg03}, which introduces a substantial simplification. An equation of state (EOS) $P(\rho)$ as well as a rotation profile (through $\Phi$) are to be prescribed in addition. Part of the technical difficulty in solving this coupled equations comes from the fact that neither the fluid boundary nor the mass density are not known in advance. A traditionnal way to capture the solutions is the Self-Consistent-Field (SCF) method, which iterates on the enthalpy field from an initial seed until convergence \citep{om68,hachisu86}. Note that this method is also used in \cite{erihach83crrr}, while \cite{ansorg03} can directly solve, without any iteration, a linear system where the unknowns are the coefficients describing the fluid boundary. The case of multi-body configurations, i.e. systems consisting in several detached bodies in mutual interactions, introduces new difficulties. There are more degrees of freedom \citep{hachisu86III}. For $m$ bodies, we have typically $2m-1$ parameters, which are for instance $m$ axis ratios and $m-1$ orbital separations. More options are possible, since each body can also have its own EOS and rotation profile \citep{hen86}.

\begin{figure}
\includegraphics[width=8.5cm,bb=0 0 656 259,clip==]{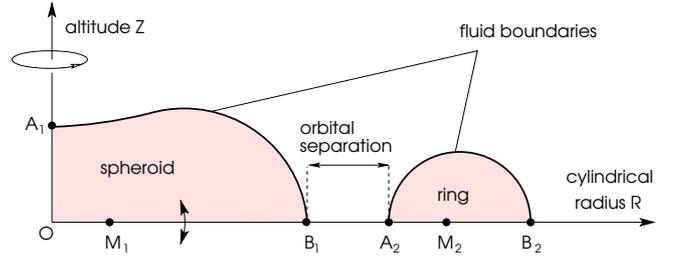}
\caption{Configuration for a spheroid-ring system. The relative separation is $s=$B$_1$A$_2/$OB$_1$ and the axis ratios are $e_i=$OA$_i/$OB$_i$ ($i=1,2$). Points M refer to the maximum enthalpy.}
\label{fig:abm.eps}
\end{figure}

The results reported here are based upon the {\tt DROP}-code whose capabibities are described in \cite{hh17}. A new version allowing to compute the equilibrium of concentric $m$-bodies in mutual interaction is fully operational. We focus on the case with $m=2$ where a ring (body number 2, or the secondary) surrounds a central body (body number 1, or the primary). The system is depicted in Fig. \ref{fig:abm.eps}. The problem is scale free. We work in dimensionless cylindrical coordinates $(\hbr,\hbz)$. The spheroid and the ring are both homogeneous (zero polytropic index) and they share the same rigid rotation law (i.e. $\hphi=-\frac{1}{2}\hbr^2$). An important assumption concerns the mass density contrast which scales the density of each body with respect to its neighbour (or to a reference body), namely
  \begin{equation}
 \eta = \rho_2/\rho_1.
  \end{equation}
This quantity is set to unity in the following, as in \cite{erihach83crrr}. In practicality, each body has its own computational grid, which is nominal in terms of accuracy and stability of the SCF method. The two Poisson equations are discretized at second-order and solved by using the multigrid method \citep{Briggs00}. Neumann Boundary Conditions (BCs) are implemented at $\hbz=0$ for the ring, and additionnally at $\hbr=0$ for the central body, and Dirichlet BCs elsewhere. All volume integrals (BCs and output quantities) take into account the fluid boundaries, which are detected through a 8-point directionnal Freeman chain code from the enthalpy maps. This is essential to get clean solutions, especially for ``hard'' EOS where $\partial \ln P/\partial \ln \rho < 1$.

\begin{figure}
  \centering
  \includegraphics[width=6.5cm,bb=142 47 651 525,clip==]{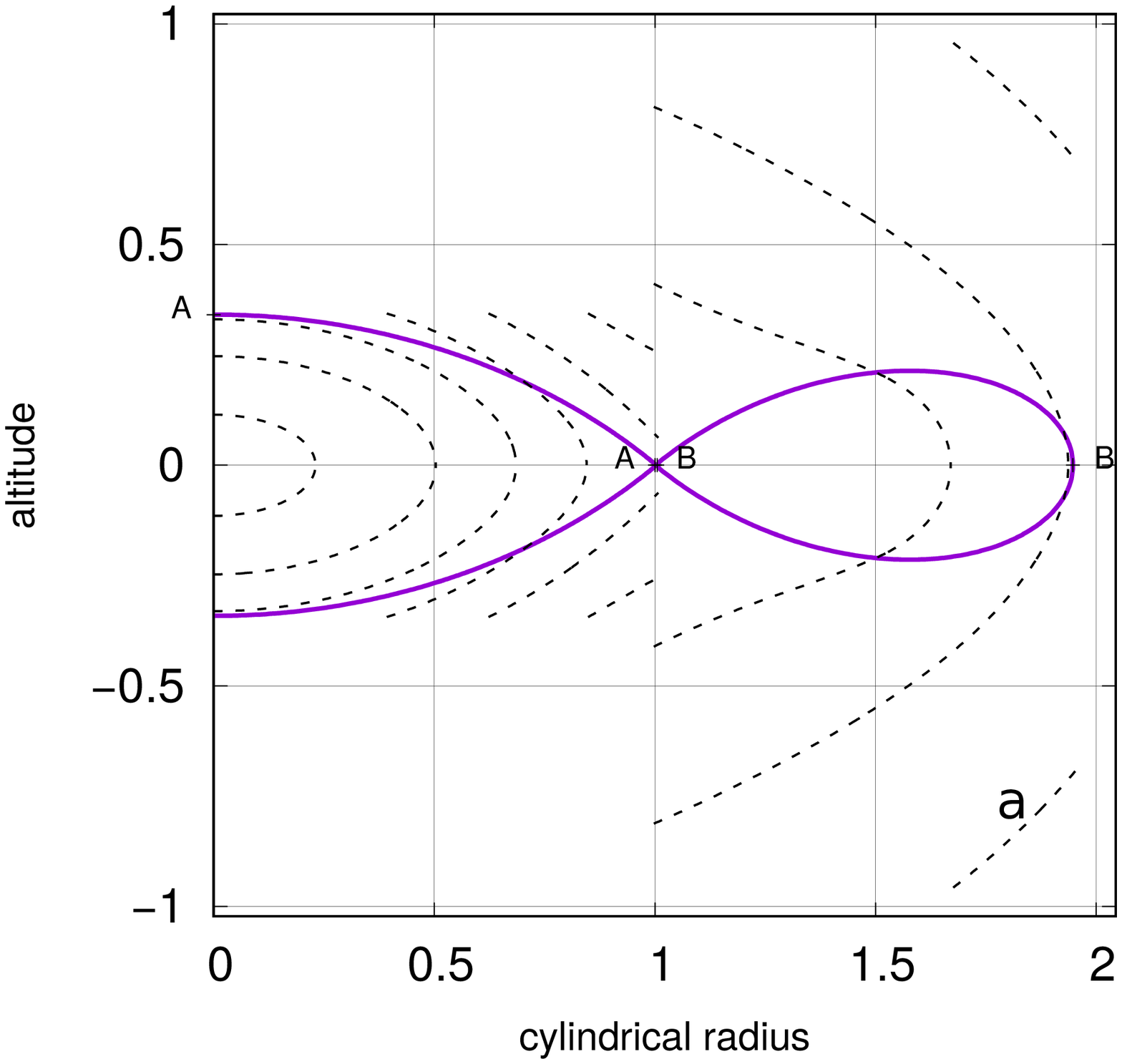}
\caption{Equilibrium state at the end-point of the $\epsilon_2$-sequence to be compared with \citep{ansorg03}. Colors stand for the fluid boundary where the enthalpy vanishes ({\it bold pink}), and a few isopotential lines are given ({\it dashed black}); see also Fig. \ref{fig:abm.eps} and Tab. \ref{tab:eps2}.}
\label{fig:epsi2.eps}
\end{figure}

If the mass density contrast is fixed and the rotation law is the same for both components, then there are only $3$ parameters for $m=2$. The most natural triplet is composed of the two axis ratios $e_i=$OA$_i/$OB$_i$ and the relative orbital separation B$_2$A$_1/$OB$_1 \equiv s$ (see Fig. \ref{fig:abm.eps}). With such a choice, however, the mass density contrast between the two bodies is not under control. We can easily impose a specific value if $\eta$ is part of the input set. We then take ($\eta=1$,$e_2$,$s$) as parameters, which means that the axis ratio of the central body $e_1$ becomes an output quantity. As observed, this does not alter the convergence properties of the SCF method. The code has been extensively checked for $m=1$, and more recently for $m$ up to $7$ \citep{bbbjmh18}. The computations have been performed with $N=129$ grid nodes per direction, which corresponds to $\ell =7$ levels of multigrid. Numerical schemes being second-order accurate in the grid spacing, errors (including the Virial parameter) are of the order of $1/N^2 \sim 10^{-4}$ typically (the results are given with four significant digits at most).

\begin{table*}
\centering
\begin{tabular}{cccccccc}\\
 \multicolumn{2}{c}{input parameters}\\
  $e_2$    & $s$  &  $M$   & $M_1/M_2$ & $j^2$ & $\omega^2$ & $e_1$ &  comment/ref. \\ \hline
  $0.5162^*$ & $0.0000$ & &$0.4105$ & $0.02972$ & $0.05433$ & $0.3413$ & \cite{ansorg03}\\ 
  $0.5162^*$ & $0.0051$ & 8.8511 &$0.4095$ & $0.02973$ & $0.05419$ & $0.3413$ & configuration a, Fig. \ref{fig:epsi2.eps}\\\hline
  $0.9900$ & $1.1204$ & 3.6708 & $1026.8$ & $0.00088$ & $0.03107$ & $0.8807$ &  configuration b, Fig. \ref{fig:eta1_youarehere.eps} \\ 
  $0.6380$ & $0.3516$ & 4.7567 &$1.0632$ & $0.02056$ & $0.05470$ & $0.6092$ & configuration c, Fig. \ref{fig:eta1_youarehere.eps}\\
  $0.9000$ & $1.9118$ & 5.3008 &$2.7043$ & $0.01148$ & $0.01318$ & $0.9304$ & configuration d, Fig. \ref{fig:eta1_youarehere.eps}\\
  $0.6000$ & $5.0000$ & 5.3798 &$0.0042$ & $0.03109$ & $0.03055$ & $0.5548$ & configuration e, Fig. \ref{fig:eta1_youarehere.eps}\\
  $0.7100$ & $1.2021$ & 11.558 &$0.3472$ & $0.03257$ & $0.02883$ & $0.7240$ & configuration f, Fig. \ref{fig:eta1_youarehere.eps}\\
  $0.6200$ & $3.0844$ & 143.49 &$0.0174$ & $0.03260$ & $0.02881$ & $0.5926$ & configuration g, Fig. \ref{fig:eta1_youarehere.eps}\\
  $0.9990$ & $0.6910$ & 3.2874 &$3\times10^5$ & $0.00170$ & $0.05306$ & $0.7906$ & critical point C, \ref{fig:etaisone.ps}\\\hline
\end{tabular}
\caption{Values obtained for the equilibria shown in Figs. \ref{fig:epsi2.eps} and \ref{fig:eta1_youarehere.eps} (see text for the numerical setup); $^*$end-point of the $\epsilon_2$-sequence. Columns 1 and 2 are input parameters. See also Fig. \ref{fig:etaisone.ps}.}
\label{tab:eps2}
\end{table*}

\begin{figure*}
  \includegraphics[width=5.8cm,bb=143 73 651 524,clip==]{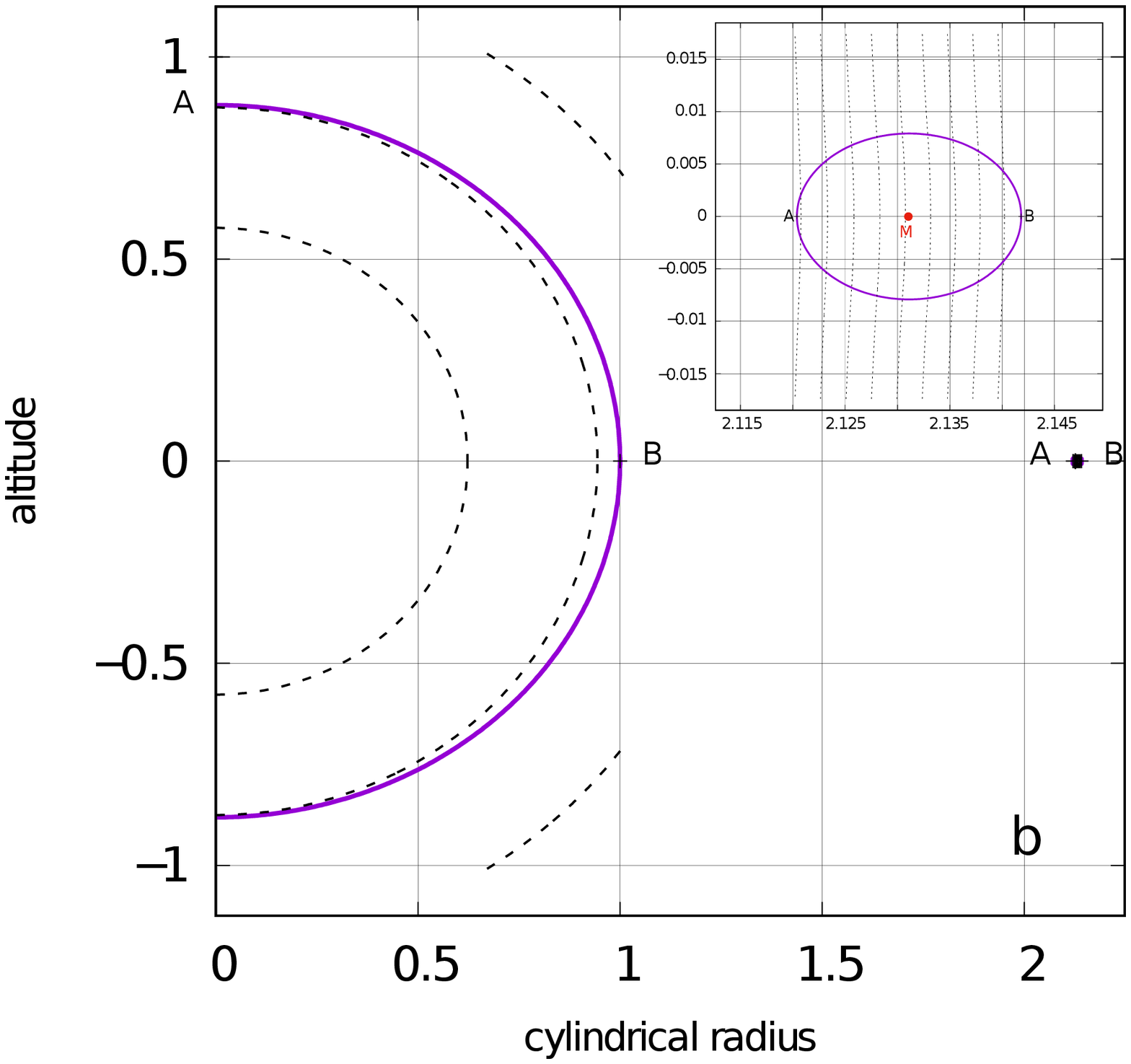}
  \includegraphics[width=5.8cm,bb=143 73 651 524,clip==]{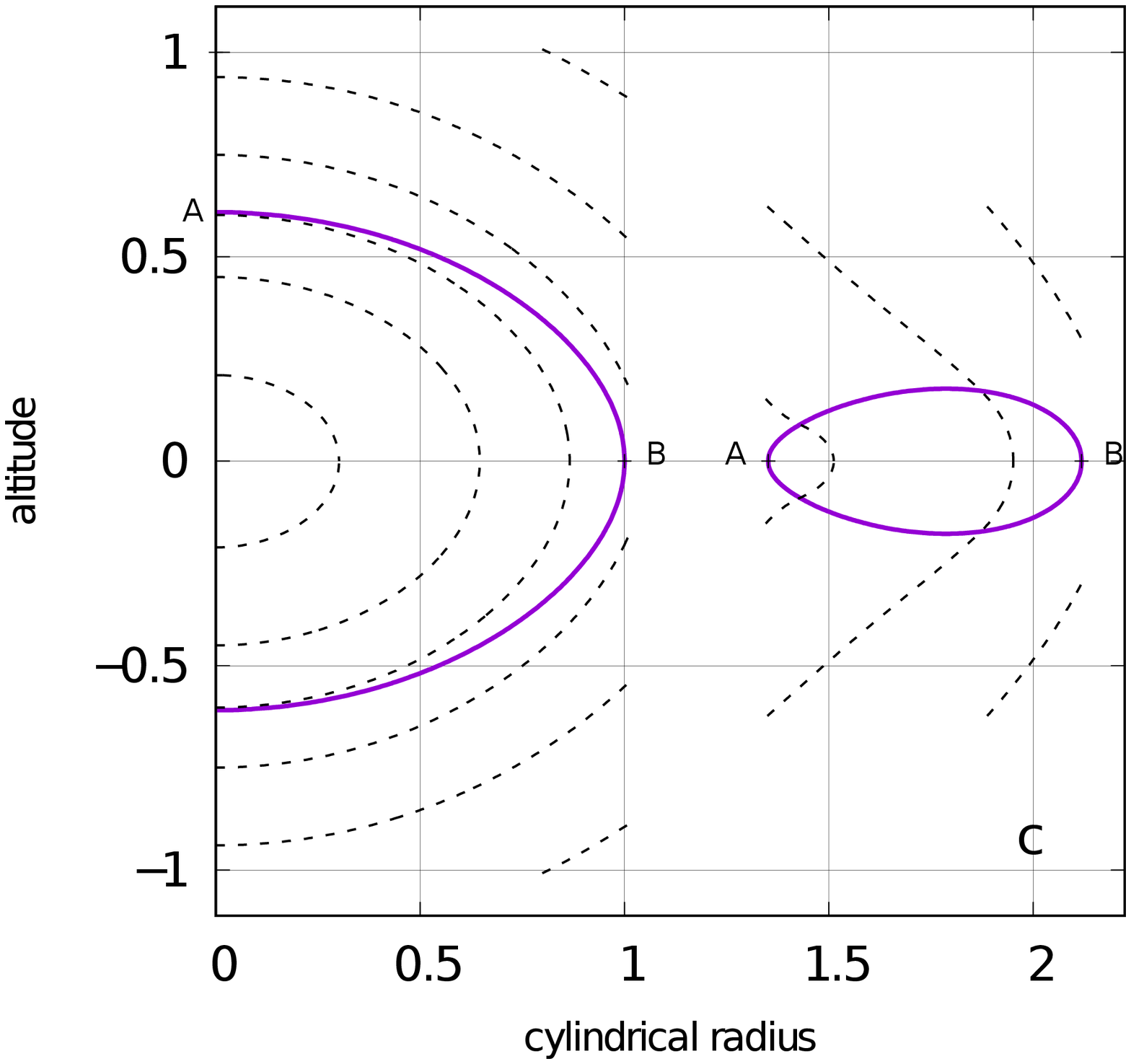}
  \includegraphics[width=5.8cm,bb=143 73 651 525,clip==]{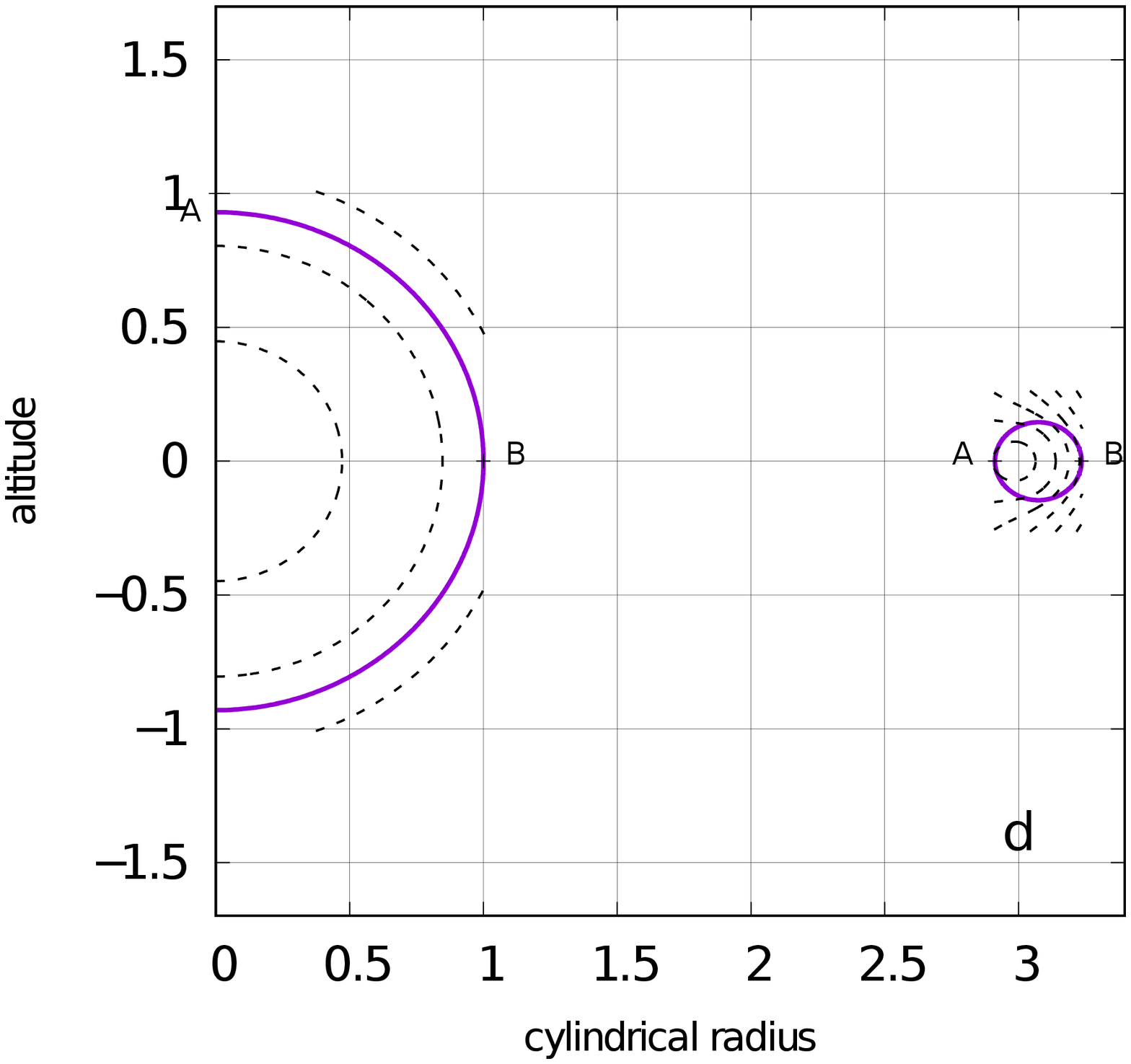}\\
  \includegraphics[width=5.8cm,bb=143 46 651 524,clip==]{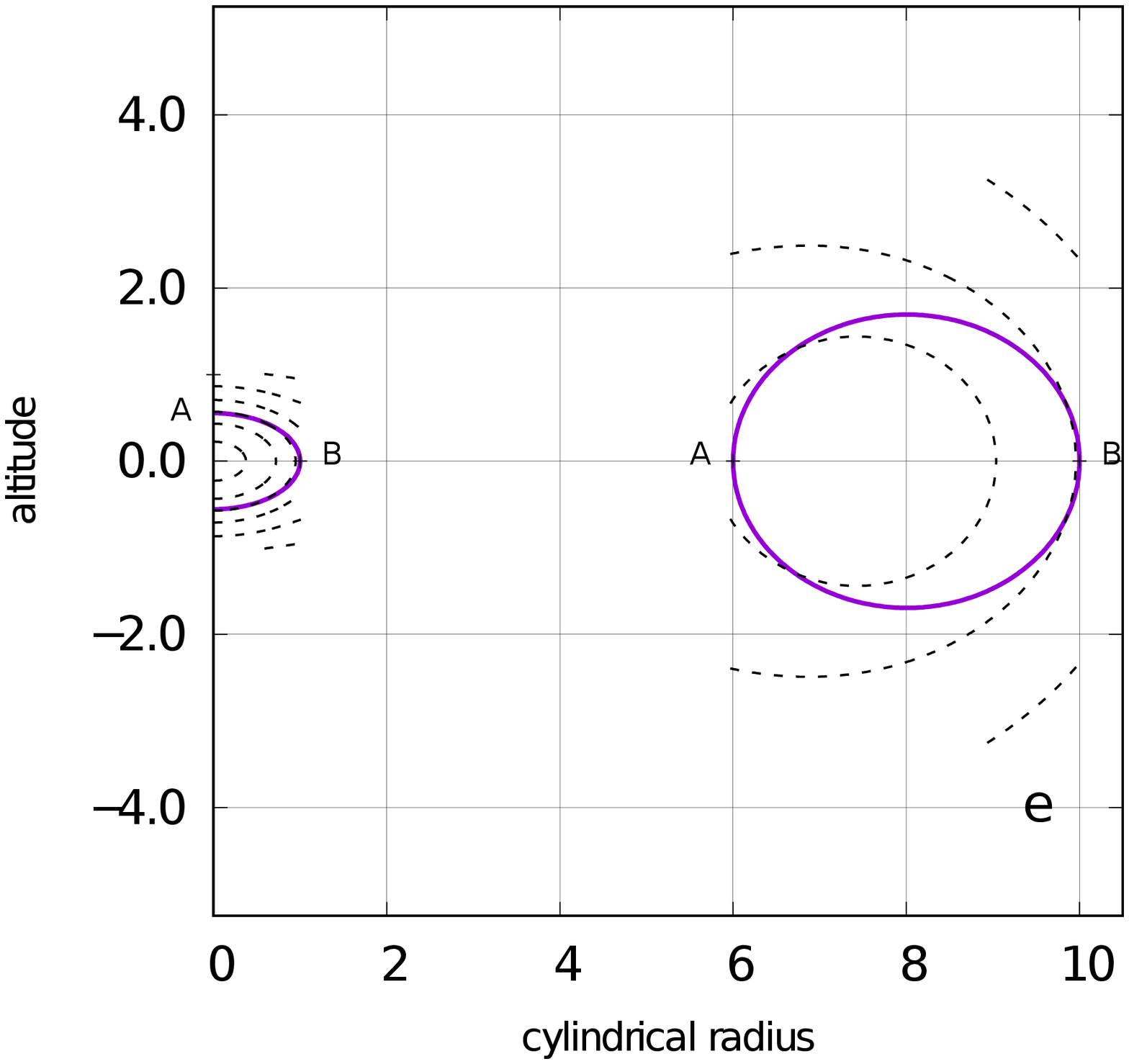}
  \includegraphics[width=5.8cm,bb=143 46 651 524,clip==]{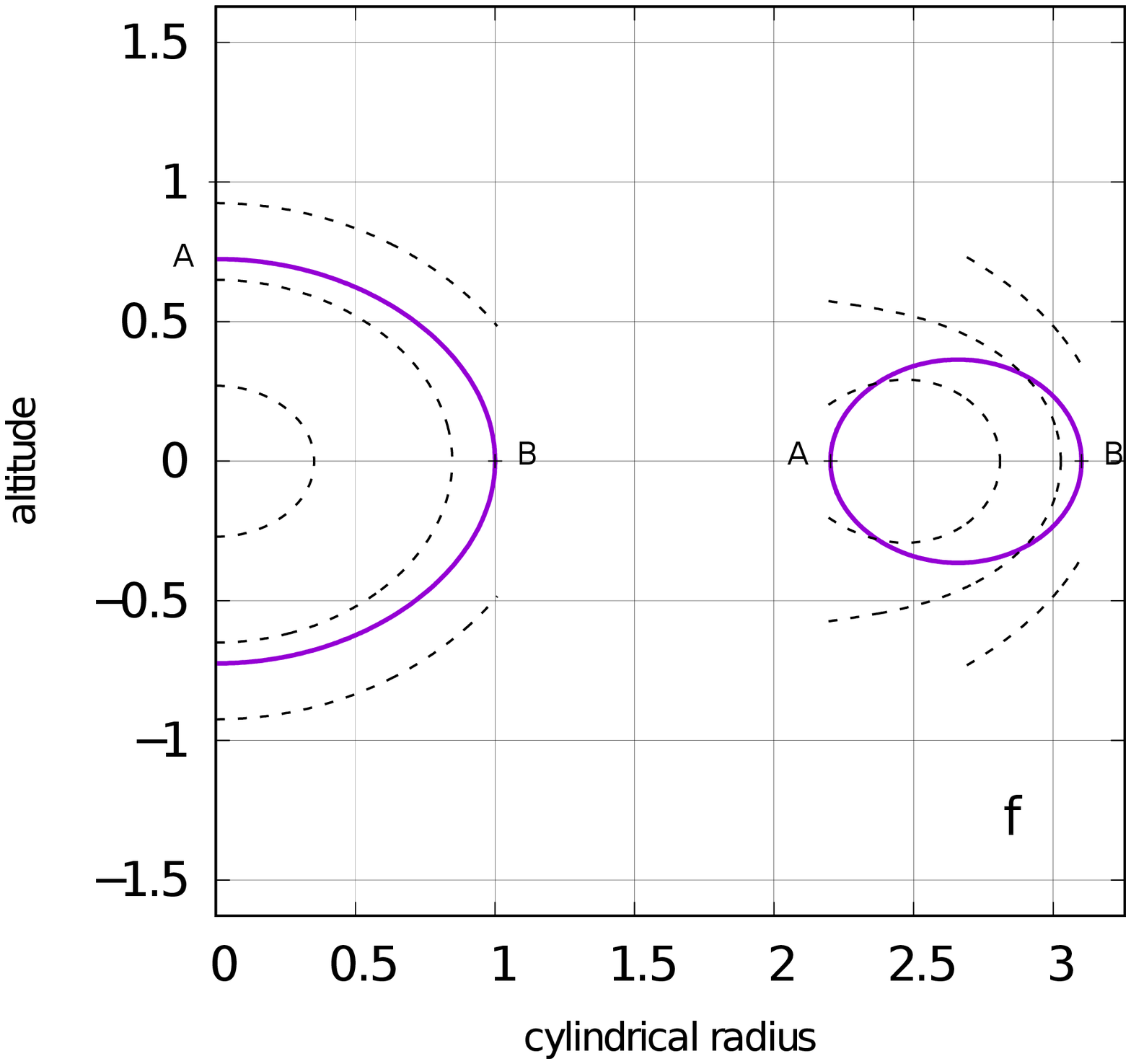}
  \includegraphics[width=5.8cm,bb=143 46 651 524,clip==]{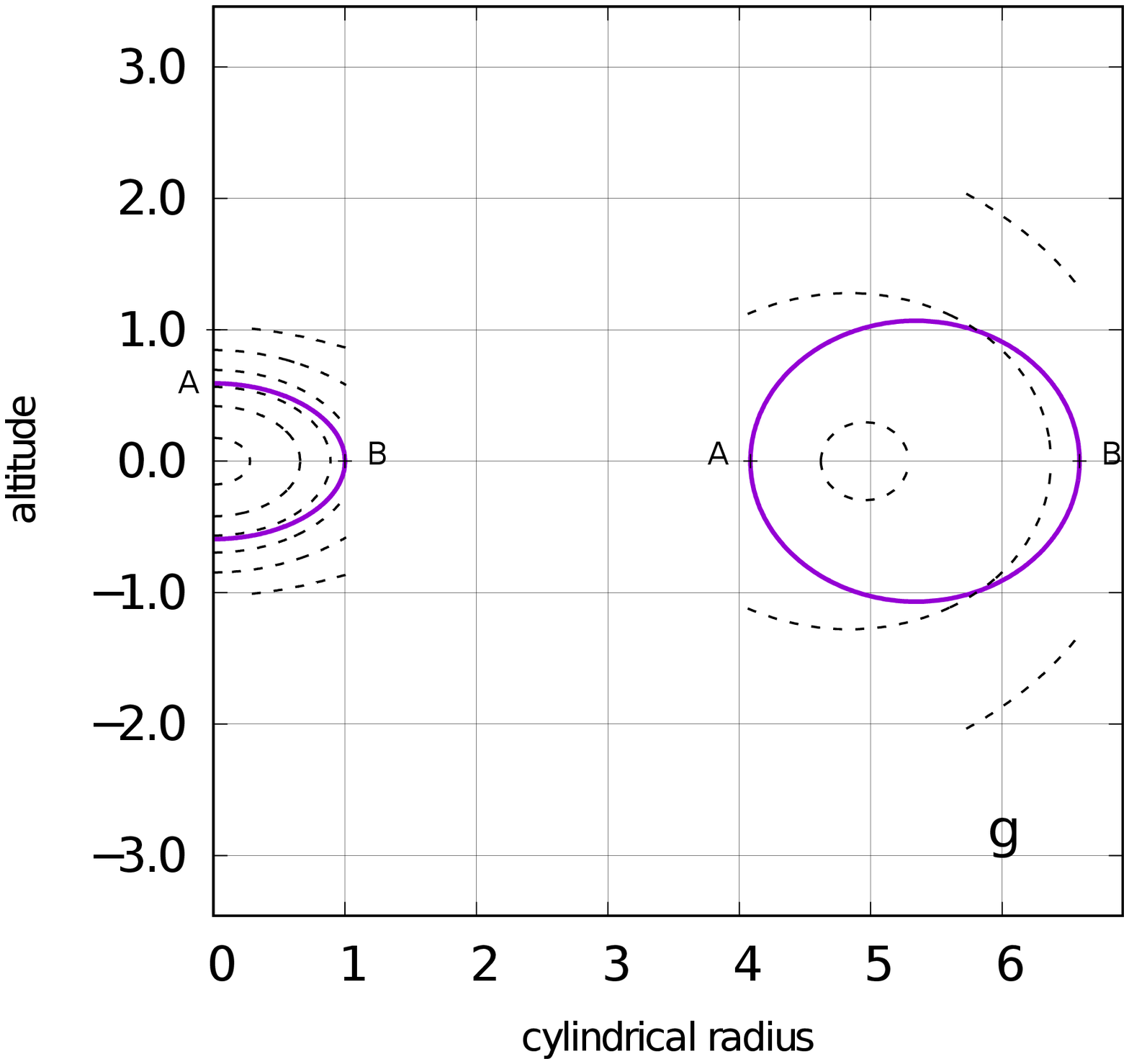}
\caption{Same legend as for Fig. \ref{fig:epsi2.eps} but for $6$ equilibria among the many runs performed (see Tab. \ref{tab:eps2} for associated key quantities) For panel b, the details of ring structure are given in the inset. }
\label{fig:eta1_youarehere.eps}
\end{figure*}

\section{Survey of the parameter space}

We have first detected the end-point of the $\epsilon_2$ sequence computed in \cite{ansorg03}. The equilibrium is shown in Fig. \ref{fig:epsi2.eps} and the output data are gathered in Tab. \ref{tab:eps2} (rows 1 and 2). We see that our results are in excellent agreement, given the adopted resolution. Next, we have generated a large number of equilibria by varying the axis ratio $e_2$ of the ring and the separation $s$ (while holding $\eta=1$). In contrast with single-body equilibria, {\it any pair of parameters does not necessarily lead to a solution}. The number of successful runs is of the order of $33000$. All computed solutions are such that $e_1 \gtrsim 0.33$ and $e_2 \gtrsim 0.51$. There is no limit in the mass ratio $M_1/M_2 \in [0,\infty[$. Besides, the orbital separation $s$ ranges from $0$ to $\infty$. A few examples of converged structures are given in Fig. \ref{fig:eta1_youarehere.eps}. Associated data are listed in Tab. \ref{tab:eps2} (rows 3 to 9). Unsurprisingly, binarity induces deviations in shape with respect to single body figures. The central body shows a slight contraction between the pole and the equator, while the ring is substantially flatter. The volume of each fluid is therefore reduced compared to the single body case. All bodies have a convex shape. We find only {\it detached binaries} (see Sect. \ref{sec:conclusion}). This means that the only system in contact system is the end-point of the $\epsilon_2$-sequence \citep{ansorg03}.

\begin{figure}
\includegraphics[width=8.5cm,bb=0 0 606 792,clip==]{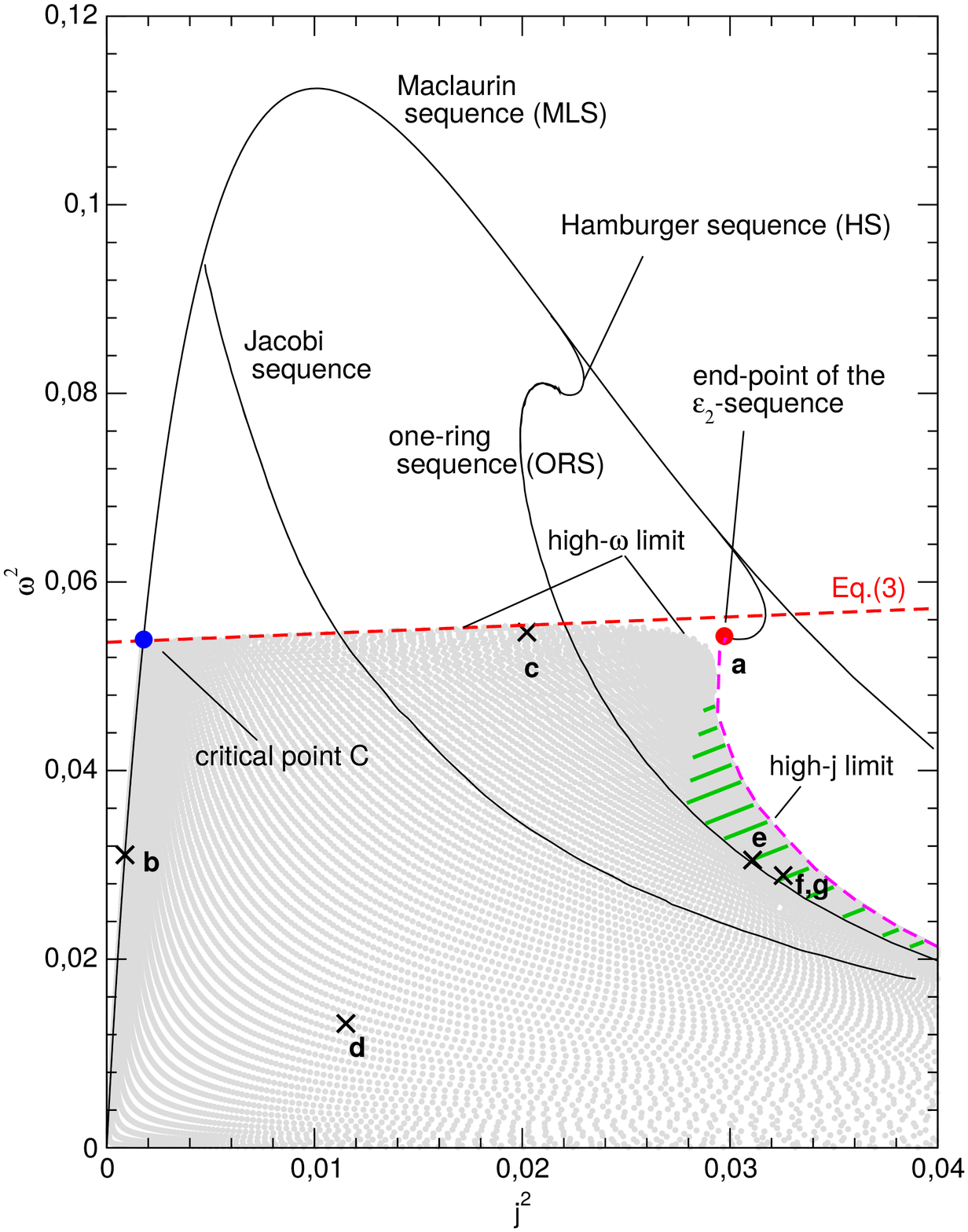}
\caption{The spheroid-ring solutions ({\it gray dots}) populate the  $\omega^2$-$j^2$ diagram in between the MLS, the high-$\omega$ limit and the high-$j$ limit (see text for explanations). The MLS, ORS, Jacobi-sequence, Hamburger-sequence and $\epsilon_2$ sequence are also shown ({\it plain lines}). Points labelled a to f ({\it cross}) correspond to equilibria shown in Fig. \ref{fig:eta1_youarehere.eps}; see also Tab. \ref{tab:eps2}. There is a band of degeneracy rightward to the ORS ({\it green dashed zone}).}
\label{fig:etaisone.ps}
\end{figure}

Figure \ref{fig:etaisone.ps} shows the results plotted in the $\omega^2$-$j^2$ diagram, where $\omega$ and $j$ are respectively the dimensionless rotation rate $\Omega$ and the dimensionless angular momentum $J$ defined by
\begin{flalign}
  \begin{cases}
      \omega^2=\frac{\Omega^2}{4\pi G \rho},\\
      j^2=\frac{J^2}{4\pi G M^{10/3} \rho^{-1/3}},
      \end{cases}
      \end{flalign}
where $M=M_1+M_2$ is the total mass. Note that there is no ambiguity in the definition of $\rho$ since the mass density contrast is unity. We have superimposed the MLS, the ORS and the Jacobi sequence for convenience. We clearly see that all solutions are concentrated between the ascending part of the MLS and a {\it first limit curve} (hereafter, the ``high-$\omega$ limit''), which meets the MLS at the critical point C $(0.00171,0.05306)$ where the axis ratio of the primary is $e_1 \approx 0.791$ and the orbital separation is $s \approx 0.691$. There is {\it no solution for $\omega \gtrsim \omega_l(j)$}. An analog is known for binary stars \citep{he84}. In the left part of it, this limit curve is close to linear, with
\begin{equation}
  \omega_l^2 \approx 0.0536 + 0.09j^2.
  \label{eq:omegal}
\end{equation}
The right-part of it is slightly bended and meets the end-point of the $\epsilon_2$-branch where $e_1 \approx 0.341$. This is due to the critical rotation (see Fig. \ref{fig:eta1_youarehere.eps}c). The configurations located close to the MLS consist in a prominent central body and a small ring (Fig. \ref{fig:eta1_youarehere.eps}b). The orbital separations are moderate to large, but non-zero (see below). For points located well in between the MLS and the ORS (Fig. \ref{fig:eta1_youarehere.eps}d), the two bodies are comparable in size and mass. The solutions overtake the ORS, and reach a {\it second limit curve} (hereafter, the ``high-$j$ limit''), which asymptotically merges with the ORS for large values of $j$, and attains the end-point of the $\epsilon_2$-sequence where $j^2 \sim 0.03$. There is probably a slight abuse of convention here in naming this limit since it does not correspond to a unique, impassable value of $j$ (this remark holds for the high-$\omega$ limit). We simply mean that, for a given value of $\omega$, there is a maximal allowed value for $j$ (with no equilibrium beyond). These maximal values tend to $\infty$ when $\omega \rightarrow 0$. Another interesting point is the presence of {\it a zone of degeneracy} located between the ORS and the high $j$-limit (Fig. \ref{fig:eta1_youarehere.eps}f and g), where two different configurations correspond to a single pair $(j^2,\omega^2)$. A point located close to the ORS can correspond to two very different configurations. Either the spheroid has a small size and relative mass, the ring dominates and the orbital separation $s$ is large (Fig. \ref{fig:eta1_youarehere.eps}e and g). Then the ring resembles very much the single ring. Or the spheroid and the ring have comparable mass and size. In this case, the ring (in particular, its axis ratio) is very different from the single ring equilibrium (Fig. \ref{fig:eta1_youarehere.eps}f). In this region where the rings are among the largest in size, the convergence of the SCF method is tricky (the number of SCF iterations rises, and the Virial parameter deteriorates).

\section{Equilibrium sequences}
\subsection{Sequences with a variable orbital separation (growth of the primary)}

\begin{figure}
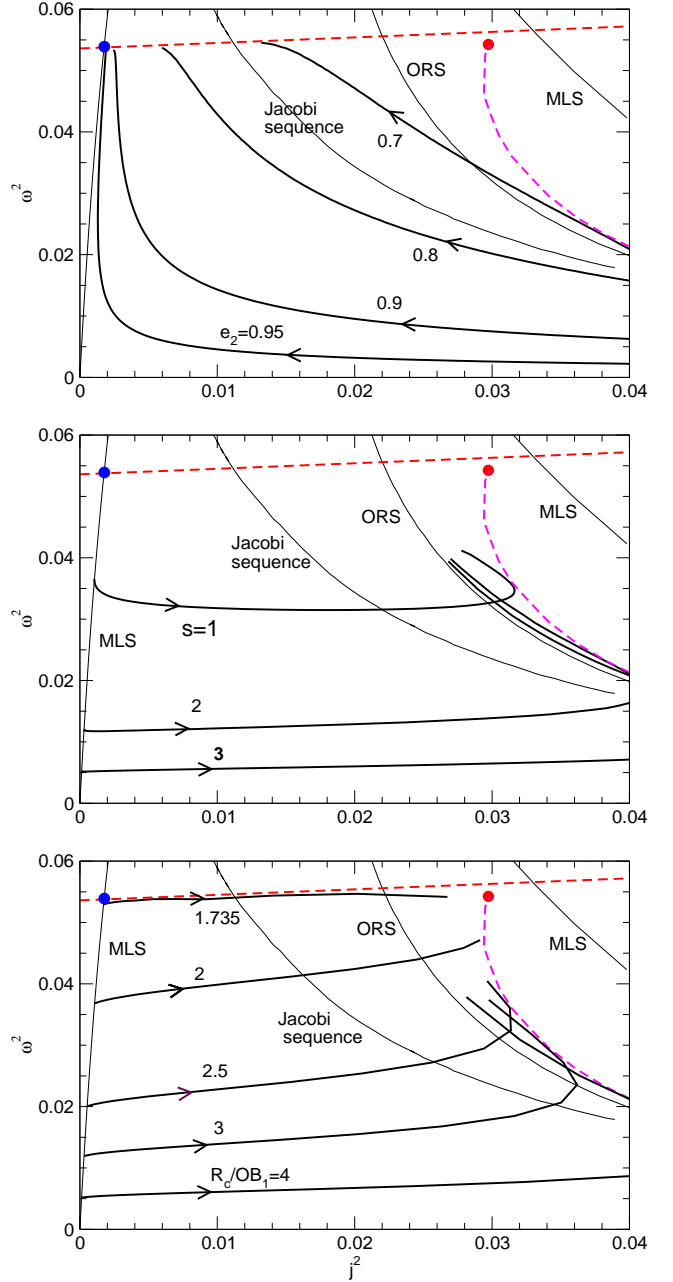

  \includegraphics[width=8.4cm,bb=4 44 601 418,clip==]{sequence_t.eps}\\

  \includegraphics[width=8.4cm,bb=4 44 601 418,clip==]{sequence_s.eps}\\

 \includegraphics[width=8.4cm,bb=4 15 601 418,clip==]{sequence_rc.eps}
\caption{Sequences obtained for constant values of $e_2$ when leaving the ORS ({\it ton panel}), for constant orbital separation $s$ ({\it middle panel}) and constant position of the ring center relative to the spheroid ({\it bottom panel}) when bifurcating from the ORS; see also Fig. \ref{fig:etaisone.ps}. Curves are labelled with the actual values of the fixed parameter.}
\label{fig:sequence_tsrc.eps}
\end{figure}

The top panel of Fig. \ref{fig:sequence_tsrc.eps} shows four sequences bifurcating from the ORS and corresponding to a decreasing orbital separation $s$, while the axis ratio of the secondary $e_2$ is held fixed. Since $s$ is infinite onto the ORS, leaving this branch means the birth and growth of a central spheroid and the decrease of the diameter of the ring. We see that, whatever the axis ratio of the initial ring on the ORS, one {\it can never reach} the MLS, but the high-$\omega$ limit where $j \gtrsim 0.00171$. When $e_2$ is significantly lower than unity, one reaches the high-$\omega$ limit more directly. When $e_2 \rightarrow 1$, the sequence is first quasi-horizontal, and then goes vertically along the MLS to finally attain the high $\omega$-limit. Asymptotically, for $e_2 \rightarrow 1$, one leaves the ORS at infinity and one reaches the critical point C where $s \approx 0.691$ (see below). As $s$ diminishes, we observe that $e_1$ is decreasing as well, which means that the central body flattens as the ring comes closer. The total mass $M$ decreases along the sequence ($j$ decreases) whereas the mass ratio $M_1/M_2$ increases.

\subsection{Sequences with a variable axis ratio for the ring (growth of the equatorial ring)}

\begin{figure}
 \includegraphics[width=5.8cm,bb=83 20 590 792,clip==,angle=-90]{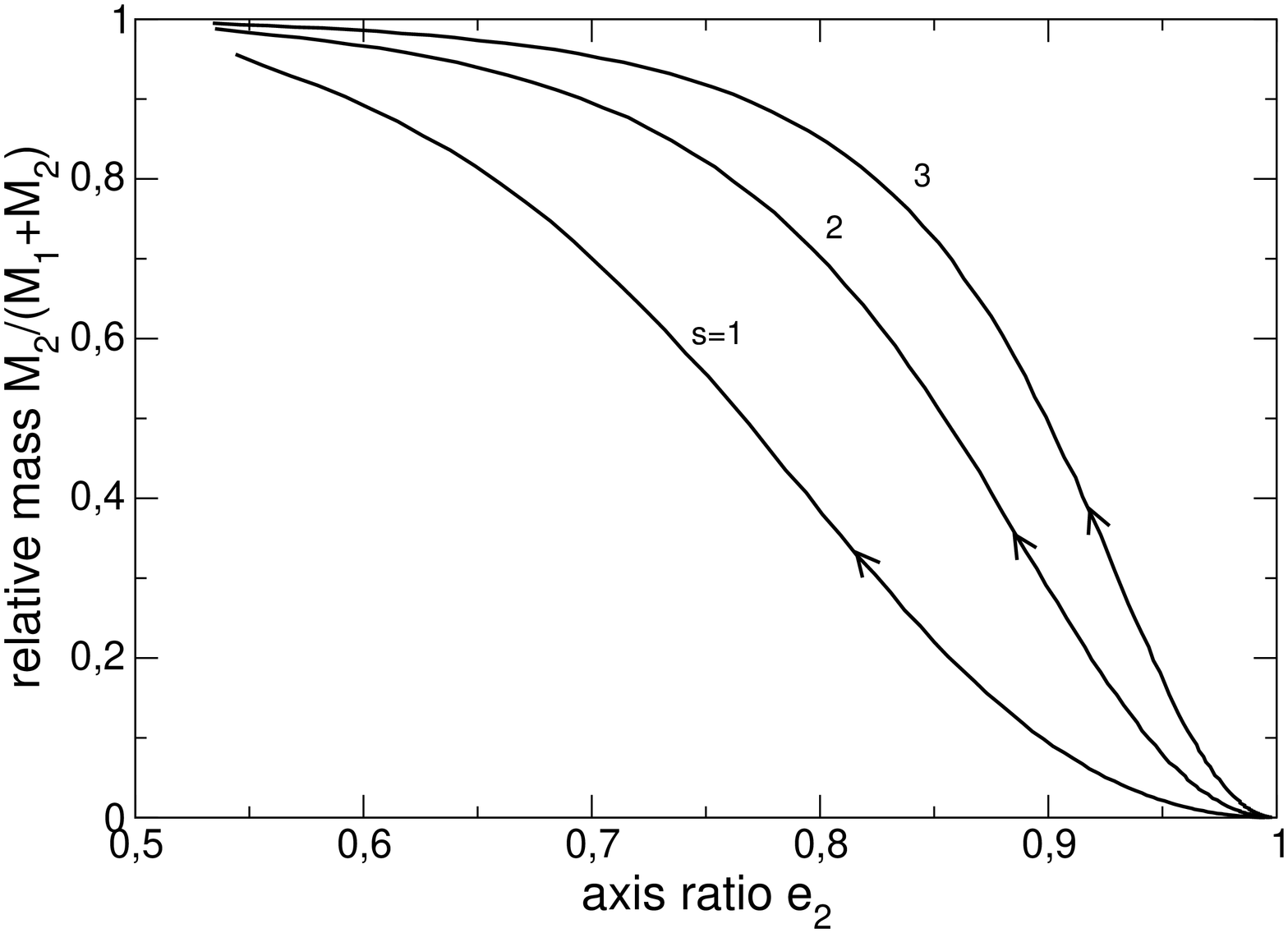}
\caption{Relative mass of the ring versus its axis ratio for the $3$ sequences obtained with a constant $s$ (see Fig. \ref{fig:sequence_tsrc.eps}, middle panel).}
\label{fig:mass_torus_s.ps}
\end{figure}

The middle panel of Fig. \ref{fig:sequence_tsrc.eps} shows three new sequences branching off from the MLS, which are obtained for a given value of the orbital separation. Leaving the MLS implies the creation and subsequent growth of an equatorial, fictitious ring with an axis ratio $e_2=1$ (i.e. a massless loop), positionned at an inital separation $s$. The lowest permitted value happens to be $\approx 0.691$. This occurs at the critical point C. Therefore, any bifurcation from the MLS means that $s$ stands in the range $[0.691,\infty[$, while $e_1 \gtrsim 0.791$. By gradually enlarging the ring, one gets closer to the ORS. We observe that the central spheroid flattens, i.e. $e_1$ decreases. The mass ratio $M_1/M_2$ is plotted versus $e_2$ in Fig. \ref{fig:mass_torus_s.ps} for these sequences. When the axis ratio of the ring is close to unity, the relative mass tends to zero and the overall equilibrium is guided by the central body (i.e. near the MLS). As soon as $e_2$ starts to decrease, the initial massless ring grows in size and in mass, and, it finally dominates the equilbrium when $e_2 \rightarrow 0.55$ (i.e. near the ORS). The total mass $M$ increases along the sequence ($j$ increases) whereas the mass ratio $M_1/M_2$ decreases.

\subsection{Sequence with fixed ring's center}

It is interesting to consider a series of solutions obtained by holding the main radius of the ring $\rc=$OA$_2+\frac{1}{2}$A$_2$B$_2$ fixed relative to the size of the primary. The bottom panel of Fig. \ref{fig:sequence_tsrc.eps} shows the results obtained for five values of $\rc/$OB$_1$. As above, we start from the MLS by enlarging an initial massless loop (a fictitious ring with zero diameter) at a relative separation $s=\rc/$OB$_1-1$. Again, bifurcations from the MLS are possible as long as we stay below the critical point C. As the ring grows, the axis ratio of the primary $e_1$ decreases as one goes towards the ORS. For the largest values of $\rc/$OB$_1$, the sequences have a large amplitude, cross the diagram rather horizontally ($\omega \sim$ const.), then overtake the ORS sequence, and go inside the band of degeneracy. These sequences end when $e_2$ reaches the nominal value of $ \approx 0.55$. The case with $\rc/$OB$_1=1.735$ is remarkable as it almost coincides with the high-$\omega$ limit. The run of $M$ and $M_1/M_2$ is the same as for the case with $s={\rm const}$.

  \subsection{Sequences with a constant mass ratio $M_1/M_2$}
   
As in \cite{erihach83crrr}, we have built sequences for which the mass ratio $M_1/M_2$ is held constant. This corresponds to systems undergoing mass exchange between the two components, or even mass-loss or accretion from the ambient medium. We can easily extract from the data cube a subset of solutions corresponding to a given output quantity $x \pm \Delta x$, where the bandwith $\Delta x$ depends on the density of the data cube (the parameter survey has not an infinite numerical resolution in $s$ and $e_2$). Sequences obtained for $M_1/M_2 \in \{0.01,0.5,1,2,10,100\}$ with an error on $M_1/M_2$ of about $\approx 10^{-4}$ typically are shown in Fig. \ref{fig:mtot} (top panel). When $M_1/M_2 \rightarrow  0$, the sequence is located near the ORS. This is expected since the equilibrium is mostly dominated by the ring. On the contrary, when $M_1/M_2\rightarrow \infty$, the sequence is close to the MLS, and it terminates on the high-$\omega$ limit. This is in agreement with \cite{erihach83crrr}. In all cases, we observe that the axis ratios $e_1$ and $e_2$ and the separation $s$ are gradually decreasing while $\omega$ increases.
    
\subsection{Sequences with a constant total mass}
    
\begin{figure}
  \includegraphics[width=8.4cm,bb=4 44 601 418,clip==]{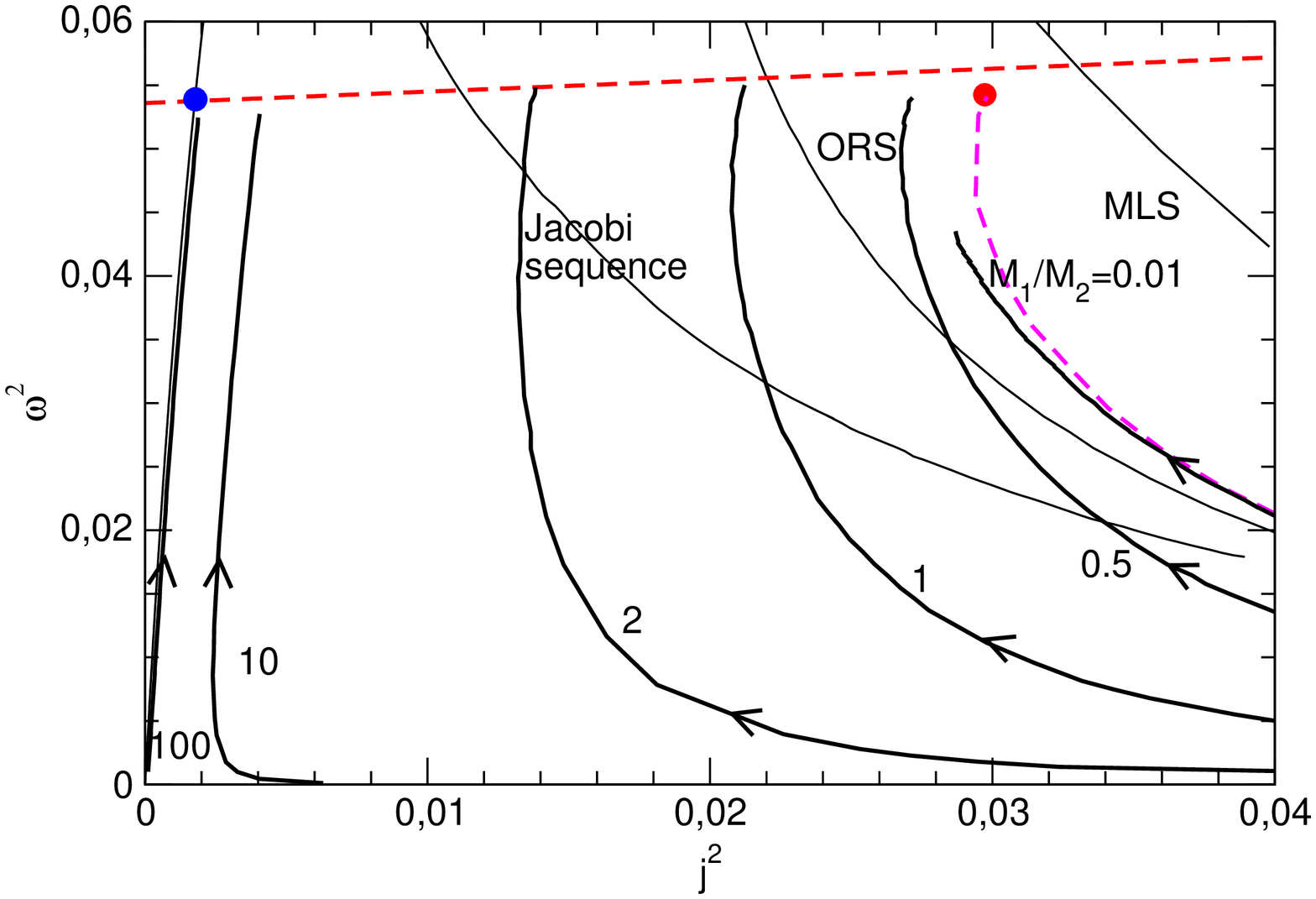}\\
  
 \includegraphics[width=8.4cm,bb=4 15 601 418,clip==]{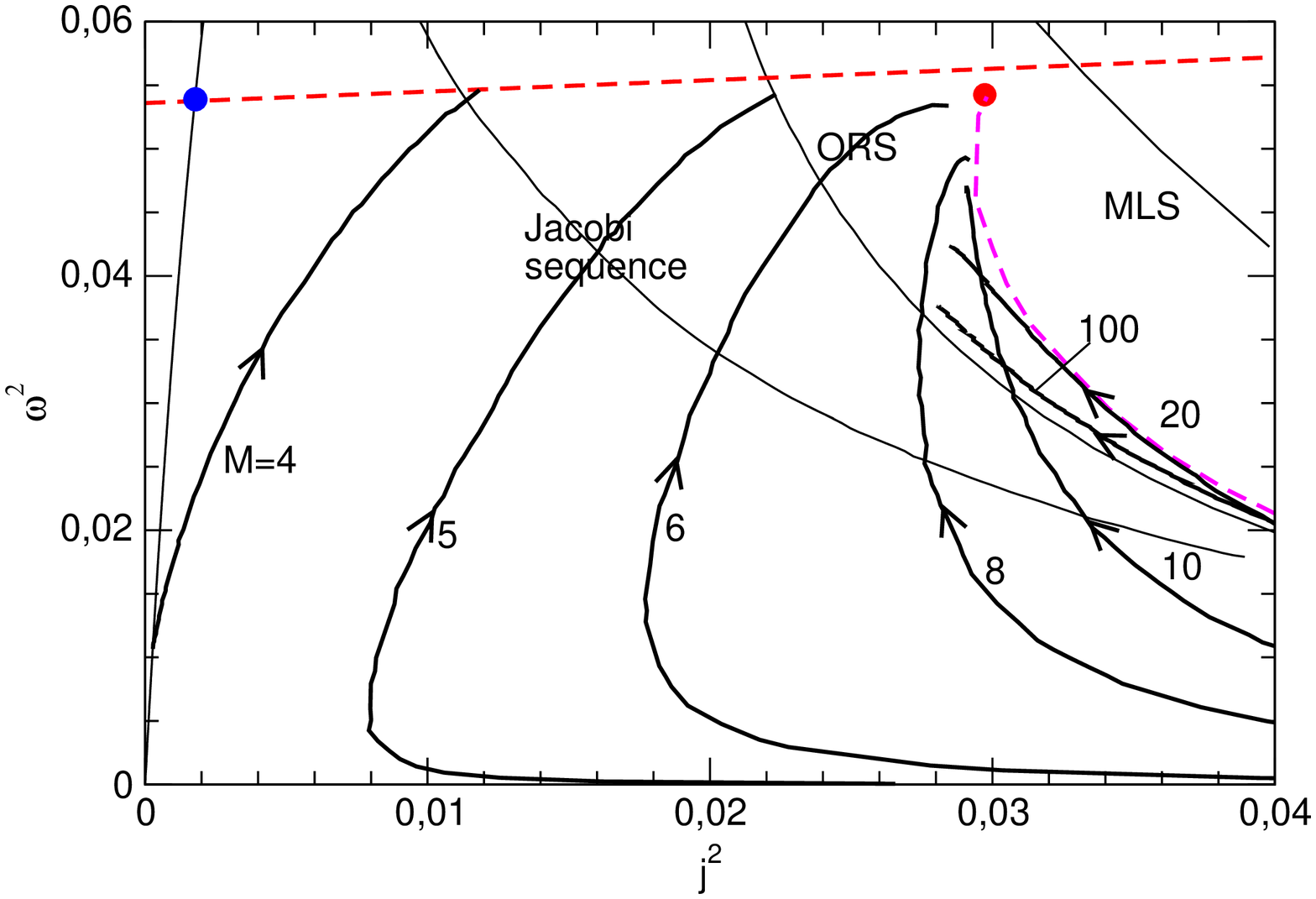}
\caption{Sequences obtained for constant values of the mass ratio $M_1/M_2$ \textit{(top panel)}, and for constant values of the total mass $M$ \textit{(bottom panel)}. For a given total mass, some sequences seem to converge towards the end-point of the $\epsilon_2$-sequence.}
\label{fig:mtot}
\end{figure}

Sequences associated with a constant total mass $M=M_1+M_2$ are also of particular interest, since a spheroid-ring system may evolve over some period with a given amount of matter, without any contact with the environment. Sequences obtained for $M= \{4,5,6,8,10.20,100\}$ (still with a precision of the order of $10^{-4}$ in relative) are displayed in Fig. \ref{fig:mtot} (bottom panel). The sequences start next to the ORS where $M_1$ and $M_2$ are comparable and, as they approach the high-$\omega$ limit, $M_1$ decreases while $M_2$ increases. Along this sequence, values of $e_1$, $e_2$ and $s$ are again decreasing while $\omega$ increases.

\section{Conclusion and perspectives}
\label{sec:conclusion}

We have studied the figures of equilibrium for spheroid-ring systems, assuming rigid rotation and incompressibility, by surveying the parameter space $(e_2,s)$ for a unity density contrast. This work is complementary to the papers by \cite{erihach83crrr} and \cite{ansorg03}. While, for single-body equilibria, there is a solution for any axis ratio, this is no more true for a spheroid-ring system. As shown, a limited set of input parameters leads to a solution, and the rotation rate must be rather low. There is no contact binary except the end-point of the $\epsilon_2$-equilibrium. Many states of critical rotation populate the equilibrium diagram in the vicinity of this singular point, for which the SCF method struggles with converging. In this zone indeed, the configurations are highly sensitive to the input parameters. A very small change in the rotation rate or/and the angular momentum implies a drastic modification of the orbital separation  and/or axis ratio, and vice versa. This sensitivity, already pointed out in \cite{erihach83crrr}, might be of great importance regarding the stability and the evolution of the binary. The end-point of the $\epsilon_2$-equilibrium seems an ideal state for any potential exchange of matter between the two components. A perturbation in the mass and dynamical content might drive the system to a very different configuration, unless some kind of selfregulation sets in. It would be interesting to investigate this question in more detail \citep[e.g.][]{wst92,akl98,mfs10}.

\cite{eh85} have shown that the MLS bifurcates towards the ORS through an intermediate body that is concave in shape. Direct routes from the MLS to the ORS are in fact possible in the ascending part of the MLS at much lower rotation rates, provided the axis ratio of the central spheroid is greater than $\approx 0.791$. At the new bifurcation points, a massless fictitious ring takes root at a relative orbital separation $s \gtrsim 0.691$, and grows up when leaving the MLS towards the ORS, while the spheroid depreciates. This clearly indicates that tiny self-gravitating rings can orbit at some distance around massive quasi-spherical bodies, planets \citep{wt88} and stars \citep{mszk06}. This meets the fundamental question of the stability of all solutions in the data cube, which would be interesting to analyze, for instance through time-dependent simulations.

We can envisage a similar approach for each $\epsilon_{k > 2}$ sequence unveiled by \cite{ansorg03}, with multiple detached rings possibly present. In particular, the end-point of the $\epsilon_3$ sequence should lead to another kind of detached binary, i.e. a ring-ring configuration \citep{erihach83crrr}. Despite such a system is probably highly unstable, it would be interesting to see if such systems display similar features (limited domain of solutions, degenerate states, bifurcations).

Finally, two severe assumptions would be worth reconsidering. The first one concerns the mass density constrast which has been set to unity. There is no reason to believe that gravitationally interacting fluids evolve with the same mass density. How the configurations computed here and the associated sequences be modified if the condition $\eta=1$ is relaxed? Besides, we have considered only incompressible fluids, which is another strong hypothesis. It would be interesting to examine the case of ``softer'' EOSs. How are the results changed for a non-zero polytropic index? The impact of the rotation law can also be investigated.

\section*{Acknowledgements}
We thank the referee, Prof. Y. Eriguchi, for fruitful comments on this paper. Part of this work has been presented at the 689. WE-Heraeus-Seminar "Accretion in strong gravity" in Bad Honnef, Germany.
\bibliographystyle{mn2e}

\begin{thebibliography}{}

\bibitem[\protect\citeauthoryear{{Abramowicz}, {Karas} \& {Lanza}}{{Abramowicz}
  et~al.}{1998}]{akl98}
{Abramowicz} M.~A.,  {Karas} V.,    {Lanza} A.,  1998, \aap, 331, 1143

\bibitem[\protect\citeauthoryear{{Ansorg}, {Kleinw{\"a}chter} \&
  {Meinel}}{{Ansorg} et~al.}{2003}]{ansorg03}
{Ansorg} M.,  {Kleinw{\"a}chter} A.,    {Meinel} R.,  2003, \mnras, 339, 515

\bibitem[\protect\citeauthoryear{{Boutin-Basillais} \&
  {Hur{\'e}}}{{Boutin-Basillais} \& {Hur{\'e}}}{2018}]{bbbjmh18}
{Boutin-Basillais} B.,  {Hur{\'e}} J.-M.,  2018, in SF2A-2018: Proceedings of
  the Annual meeting of the French Society of Astronomy and Astrophysics
  {Multi-body figures of equilibrium in axial symmetry}.
pp 111--112

\bibitem[\protect\citeauthoryear{Briggs, Henson \& McCormick}{Briggs
  et~al.}{2000}]{Briggs00}
Briggs W.~L.,  Henson V.~E.,    McCormick S.~F.,  2000, A multigrid tutorial
  (2nd ed.).
Society for Industrial and Applied Mathematics, Philadelphia, PA, USA

\bibitem[\protect\citeauthoryear{{Chandrasekhar}}{{Chandrasekhar}}{1969}]{chandra69}
{Chandrasekhar} S.,  1969, {Ellipsoidal figures of equilibrium}.
Yale Univ. Press

\bibitem[\protect\citeauthoryear{{Eriguchi} \& {Hachisu}}{{Eriguchi} \&
  {Hachisu}}{1983}]{erihach83crrr}
{Eriguchi} Y.,  {Hachisu} I.,  1983, Progress of Theoretical Physics, 69, 1131

\bibitem[\protect\citeauthoryear{{Eriguchi} \& {Hachisu}}{{Eriguchi} \&
  {Hachisu}}{1985}]{eh85}
{Eriguchi} Y.,  {Hachisu} I.,  1985, \aap, 148, 289

\bibitem[\protect\citeauthoryear{{Fujisawa} \& {Eriguchi}}{{Fujisawa} \&
  {Eriguchi}}{2014}]{fe14}
{Fujisawa} K.,  {Eriguchi} Y.,  2014, \mnras, 438, L61

\bibitem[\protect\citeauthoryear{{Hachisu}}{{Hachisu}}{1986a}]{hachisu86}
{Hachisu} I.,  1986a, \apjs, 61, 479

\bibitem[\protect\citeauthoryear{{Hachisu}}{{Hachisu}}{1986b}]{hachisu86III}
{Hachisu} I.,  1986b, \apjs, 62, 461

\bibitem[\protect\citeauthoryear{{Hachisu} \& {Eriguchi}}{{Hachisu} \&
  {Eriguchi}}{1984}]{he84}
{Hachisu} I.,  {Eriguchi} Y.,  1984, \apss, 99, 71

\bibitem[\protect\citeauthoryear{{Hachisu}, {Eriguchi} \& {Nomaoto}}{{Hachisu}
  et~al.}{1986a}]{hen86}
{Hachisu} I.,  {Eriguchi} Y.,    {Nomoto} K.,  1986a, \apj, 308, 161

\bibitem[\protect\citeauthoryear{{Hachisu}, {Eriguchi} \& {Nomoto}}{{Hachisu}
  et~al.}{1986b}]{henii86}
{Hachisu} I.,  {Eriguchi} Y.,    {Nomoto} K.,  1986b, \apj, 311, 214

\bibitem[\protect\citeauthoryear{{Horedt}}{{Horedt}}{2004}]{horedttextbook2004}
{Horedt} G.~P.,  ed. 2004, {Polytropes - Applications in Astrophysics and
  Related Fields} Vol.~306 of Astrophysics and Space Science Library

\bibitem[\protect\citeauthoryear{{Hur{\'e}} \& {Hersant}}{{Hur{\'e}} \&
  {Hersant}}{2017}]{hh17}
{Hur{\'e}} J.-M.,  {Hersant} F.,  2017, \mnras, 464, 4761

\bibitem[\protect\citeauthoryear{{Hur{\'e}}, {Hersant} \& {Nasello}}{{Hur{\'e}}
  et~al.}{2018}]{hhn18}
{Hur{\'e}} J.-M.,  {Hersant} F.,    {Nasello} G.,  2018, \mnras, 475, 63

\bibitem[\protect\citeauthoryear{{Kratter} \& {Matzner}}{{Kratter} \&
  {Matzner}}{2006}]{km06}
{Kratter} K.~M.,  {Matzner} C.~D.,  2006, \mnras, 373, 1563

\bibitem[\protect\citeauthoryear{{Lock} \& {Stewart}}{{Lock} \&
  {Stewart}}{2017}]{ls17}
{Lock} S.~J.,  {Stewart} S.~T.,  2017, Journal of Geophysical Research
  (Planets), 122, 950

\bibitem[\protect\citeauthoryear{{Mason}}{{Mason}}{2015}]{mas15}
{Mason} R.~E.,  2015, \planss, 116, 97

\bibitem[\protect\citeauthoryear{{Masuda}, {Nishida} \& {Eriguchi}}{{Masuda}
  et~al.}{1998}]{mne98}
{Masuda} N.,  {Nishida} S.,    {Eriguchi} Y.,  1998, \mnras, 297, 1139

\bibitem[\protect\citeauthoryear{{Meilland}, {Stee}, {Zorec} \&
  {Kanaan}}{{Meilland} et~al.}{2006}]{mszk06}
{Meilland} A.,  {Stee} P.,  {Zorec} J.,    {Kanaan} S.,  2006, \aap, 455, 953

\bibitem[\protect\citeauthoryear{{Montero}, {Font} \& {Shibata}}{{Montero}
  et~al.}{2010}]{mfs10}
{Montero} P.~J.,  {Font} J.~A.,    {Shibata} M.,  2010, Physical Review
  Letters, 104, 191101

\bibitem[\protect\citeauthoryear{{Nishida}, {Eriguchi} \& {Lanza}}{{Nishida}
  et~al.}{1992}]{nel92}
{Nishida} S.,  {Eriguchi} Y.,    {Lanza} A.,  1992, \apj, 401, 618

\bibitem[\protect\citeauthoryear{{Ostriker} \& {Mark}}{{Ostriker} \&
  {Mark}}{1968}]{om68}
{Ostriker} J.~P.,  {Mark} J.~W.-K.,  1968, \apj, 151, 1075

\bibitem[\protect\citeauthoryear{{Takahashi} \& {Inutsuka}}{{Takahashi} \&
  {Inutsuka}}{2016}]{ti16}
{Takahashi} S.~Z.,  {Inutsuka} S.-i.,  2016, \aj, 152, 184

\bibitem[\protect\citeauthoryear{{Trova}, {Schroven}, {Hackmann}, {Karas},
  {Kov{\'a}{\v r}} \& {Slan{\'y}}}{{Trova} et~al.}{2018}]{audrey18}
{Trova} A.,  {Schroven} K.,  {Hackmann} E.,  {Karas} V.,  {Kov{\'a}{\v r}} J.,
    {Slan{\'y}} P.,  2018, \prd, 97, 104019

\bibitem[\protect\citeauthoryear{{Wisdom} \& {Tremaine}}{{Wisdom} \&
  {Tremaine}}{1988}]{wt88}
{Wisdom} J.,  {Tremaine} S.,  1988, \aj, 95, 925

\bibitem[\protect\citeauthoryear{{Wong}}{{Wong}}{1974}]{wong74}
{Wong} C.-Y.,  1974, \apj, 190, 675

\bibitem[\protect\citeauthoryear{{Woodward}, {Sankaran} \&
  {Tohline}}{{Woodward} et~al.}{1992}]{wst92}
{Woodward} J.~W.,  {Sankaran} S.,    {Tohline} J.~E.,  1992, \apj, 394, 248

\end{thebibliography}

\end{document}